\documentclass[12pt,english,a4paper]{amsart}
\usepackage[T1]{fontenc}
\usepackage[latin1]{inputenc}
\usepackage[a4paper]{geometry}
\usepackage{amsthm}
\usepackage{amsbsy}
\usepackage{amstext}
\usepackage{graphicx}
\usepackage{amssymb}

\makeatletter
\numberwithin{equation}{section}
\numberwithin{figure}{section}
  \theoremstyle{remark}
  \newtheorem*{rem*}{Remark}

\makeatother

\usepackage{babel}

\begin{document}
\global\long\def\Cc{\mathbb{C}}

\global\long\def\Rr{\mathbb{R}}

\global\long\def\Nn{\mathbb{N}}

\global\long\def\Qq{\mathbb{Q}}

\global\long\def\Zz{\mathbb{Z}}

\global\long\def\eps{\varepsilon}

\global\long\def\borelB{\mathcal{B}}

\global\long\def\baireM{\mathcal{M}}

\global\long\def\metricd{\mathsf{d}}

\global\long\def\DStar{\mathcal{D}^{*}}

\global\long\def\mapA{\mathcal{A}}

\global\long\def\lebesgueL{\mathcal{L}}

\global\long\def\hausdorffH{\mathcal{H}}

\global\long\def\mathS{\mathcal{S}}

\global\long\def\mathP{\mathcal{P}}

\global\long\def\vaguetopology{\mathcal{T}_{\mathcal{V}}}

\global\long\def\vagueV{\mathcal{V}}

\global\long\def\support{\textnormal{Support}}

\global\long\def\muomega{\mu_{\omega}}

\global\long\def\muomegaeps{\mu_{\omega}^{\eps}}

\global\long\def\mutildeomegaeps{\mu_{\widetilde{\omega}}^{\eps}}

\global\long\def\mupalm{\mu_{\mathcal{P}}}

\global\long\def\nupalm{\nu_{\mathcal{P}}}

\global\long\def\mugammaomegaeps{\mu_{\Gamma(\omega)}^{\eps}}

\global\long\def\mugammaomega{\mu_{\Gamma(\omega)}}

\global\long\def\mugammapalm{\mu_{\Gamma,\mathcal{P}}}

\global\long\def\tauxeps{\tau_{\frac{x}{\eps}}}

\global\long\def\tildeomega{\widetilde{\omega}}

\global\long\def\wOmega{\widetilde{\Omega}}

\global\long\def\womega{\widetilde{\omega}}

\global\long\def\wsigma{\widetilde{\sigma}}

\global\long\def\wGamma{\widetilde{\Gamma}}

\global\long\def\wtau{\widetilde{\tau}}

\global\long\def\wdelta{\widetilde{\delta}}

\global\long\def\wphi{\widetilde{\phi}}

\global\long\def\wmu{\widetilde{\mu}}

\global\long\def\weta{\widetilde{\eta}}

\global\long\def\closedsets{\mathcal{F}}

\global\long\def\ttopology{\mathcal{T}_{\closedsets}}

\global\long\def\intrn{\int_{\Rr^{n}}}

\global\long\def\intomega{\int_{\Omega}}

\global\long\def\spaceomega{(\Omega,\borelB(\Omega),\mu)}

\global\long\def\sigmafinite{$\sigma$-finite~}

\global\long\def\limarrow#1#2{{\displaystyle \stackrel{\longrightarrow}{#1\rightarrow#2}}}

\global\long\def\Lpomega{L^{p}\spaceomega}

\global\long\def\nablax{\nabla_{x}}

\global\long\def\nablaomega{\nabla_{\omega}}

\renewcommand{\div}{{\text{div}\,}}

\global\long\def\divx{\textnormal{div}_{x}}

\global\long\def\divomega{\textnormal{div}_{\omega}}

\global\long\def\curlx{\curl_{x}}

\global\long\def\curlomega{\curl_{\omega}}

\global\long\def\grad{\textnormal{grad}}

\global\long\def\gradx{\grad_{x}}

\global\long\def\gradomega{\grad_{\omega}}

\global\long\def\Ltwomuomegaepsmu{L^{2}([\Omega,\mu]\,;\, L^{2}(\Rr^{n},\mu_{\omega}^{\eps}))}

\global\long\def\Ltwomupalmdx{L^{2}([\Rr^{n},\lebesgueL]\,;\, L^{2}(\Omega,\mupalm))}

\global\long\def\Ltwomuomegaeps{L^{2}(\Rr^{n},\muomegaeps)}

\global\long\def\weaktwomean{\stackrel{w2sm}{\longrightarrow}}

\global\long\def\strongtwomean{\stackrel{s2sm}{\longrightarrow}}

\global\long\def\mikelic{Mikeli\'c}

\global\long\def\cnull{C_{0}(\Rr^{n})}

\global\long\def\tautildeomegaeps{\widetilde{\tau}_{\frac{x}{\eps}\omega}}

\global\long\def\diveromega{\mbox{div}_{\omega}}

\global\long\def\Y{\boldsymbol{Y}}

\global\long\def\Q{\boldsymbol{Q}}

\global\long\def\diver{\mbox{div}\,}

\global\long\def\diverx{\mbox{div}_{x}}

\global\long\def\divery{\mbox{div}_{y}}

\global\long\def\vel{\upsilon}

\newtheorem{thm}{Theorem}
\newtheorem{proposition}[thm]{Proposition} 

\newtheorem{definition}[thm]{Definition} 
\newtheorem{theorem}[thm]{Theorem} 
\newtheorem{lemma}[thm]{Lemma} 
\newtheorem{korollary}[thm]{Korollary} 
\newtheorem{remark}[thm]{Remark} 
\newtheorem{example}[thm]{Example} 
\newtheorem{condition}[thm]{Condition} 

\title[Asymptotic Expansion on Stochastic Geometries]{Asymptotic Expansion for Multiscale Problems on Non-periodic Stochastic
Geometries}

\author{Martin Heida}

\address{Institute for Applied Mathematics\\
Im Neuenheimer Feld 294\\
D-69120 Heidelberg\\
Germany}

\email{martin.heida@iwr.uni-heidelberg.de}
\begin{abstract}
The asymptotic expansion method is generalized from the periodic setting
to stationary ergodic stochastic geometries. This will demonstrate
that results from periodic asymptotic expansion also apply to non-periodic
structures of a certain class. In particular, the article adresses
non-mathematicians who are familiar with asymptotic expansion and
aims at introducing them to stochastic homogenization in a simple
way. The basic ideas of the generalization can be formulated in simple
terms, which is basically due to recent advances in mathematical stochastic
homogenization. After a short and formal introduction of stochastic
geometry, calculations in the stochastic case will be formulated in
a way that they will not look different from the periodic setting.
To demonstrate that, the method will be applied to diffusion with
and without microscopic nonlinear boundary conditions and to porous
media flow. Some examples of stochastic geometries will be given.
\end{abstract}

\subjclass[2000]{35B27, 80M40, 74Q10, 60D05}

\keywords{stochastic homogenization, asymptotic expansion, two-scale}

\maketitle

\section{Introduction}

Homogenization has become an important modeling tool for multiscale
problems, i.e. phenomena that have causes and effects on multiple
spatial scales. The application ranges from Physics (porous media
flow, freezing processes in porous media, erosion\cite{Allaire1991,Eck2004a,Peter2009}),
engineering (composite materials, reaction diffusion equations in
catalysts\cite{Neuss-Radu2001,Hummel2000}) to biology (processes
in tissue\cite{Marciniak-Czochra2008}, membranes\cite{Neuss-Radu2008}).
Also the numerical investigation of such problems is of interest (see
\cite{Miehe2007} and references therein).

Homogenization considers physical, chemical or biological processes
on large domains with a periodic microstructure of period $\eps\ll1$.
On this periodic structure, a set of equations describing the several
processes  are set up and the limit behavior of the solutions of
these equations is investigated as $\eps\rightarrow0$. Note that
$\eps\rightarrow0$ is a mathematical equivalent of the assumption
$\eps\ll1$, which means basically very small microstructures compared
to the domain of interest. For example, one may think of a solid structure
perforated by a periodic structure of channels, which are filled by
a Newtonian fluid, i.e. a fluid whose motion is described by the Navier-Stokes
equations. Since one is usually interested in a scale of meters and
the pores have sizes of $mm$, it follows $\eps\approx10^{-3}$. In
the limit $\eps\rightarrow0$, the velocity field would follow Darcy's
law \cite{Hornung1997} (see sections \ref{sub:Porous-media-flow-per}
and \ref{sub:Porous-media-flow-sto}).

There are two ways to obtain the limit equations: Via strong mathematical
calculations (proofs of convergence of solutions) or via rather formal
calculations. One of the formal ways to obtain averaged equations
is the so called asymptotic expansion. Its ansatz is to expand the
solution in a series of functions multiplied with increasing powers
of $\eps$, where the functions in this expansion depend on a global
variable and a periodic variable. We will demonstrate the basics of
this approach below in section \ref{sec:Recapitulation-of-Periodic}.

Strong mathematical investigations where originally based on strong
and weak convergence methods in Sobolev spaces, which is nowadays
still a method of choice. However, lots of other convergence methods
entered homogenization theory such as $\Gamma$- and $G$-convergence,
see \cite{Zhikov1994} for an overview.

In the beginning of the 1990's, Allaire and Nguetseng \cite{Nguetseng1989,Allaire1992}
developed the so called two-scale convergence which was later extended
by Neuss-Radu, Zhikov and Lukkassen and Wall \cite{Neuss-Radu1996,Zhikov2000,Lukkassen2005}.
As an interesting feature, two-scale convergence is related to an
asymptotic expansion in the weak sense and therefore closely connected
to the formal method. This similarity becomes even more striking by
the method of periodic unfolding, developed by Cioranescu, Damlamian
and Griso \cite{Cioranescu2002,Cioranescu2008}, and which can be
considered as a consequent generalization of two-scale convergence.
In comparing the methods, one may get the idea that every result from
formal asymptotic expansion could be proved rigorously. However, for
many homogenization problems there are lots of technical difficulties
like regularity proofs or a lack of Poincaré-inequalities which are
sometimes to hard to be overcome.

There is, however, the justified criticism, that the methods above
are based on the assumption of a periodic structure and that the resulting
equations therefore may not be valid for a non-periodic medium. Nevertheless,
we expect that many of the derived equations may also hold for non-periodic
microscopic geometries: Darcy's law is observed to hold in most porous
media may it be sand, silt or loam. In a polycrystal or a composite
material (industrial ceramics) with their complex microstructure,
we assume that heat transport still follows Fourier's law with averaged
coefficients. Reaction diffusion equations derived for a catalyzer
should also hold if the microscopic structure is not perfectly periodic.

So, it is an important question whether or not it is possible to apply
homogenized results to non-periodic geometries and under what circumstances.
Mathematicians were aware of this problem and tried to apply homogenization
techniques to non-periodic structures. An overview over many results
before 1994 can be found in \cite{Zhikov1994}. 

In 2006, Zhikov and Piatnitsky \cite{Zhikov2006} where able to introduce
a two-scale convergence method on stationary, ergodic stochastic geometries
on a compact probability space. Their results where generalized to
arbitrary probability spaces by the author in \cite{Heida2009}. A
former attempt by Bourgeat, Mikeli\'{c} and Wright \cite{Bourgeat1994}
in the non-ergodic setting used an averaged form of two-scale convergence
and was not applicable to problems on manifolds or complex boundary
conditions. Also, due to the averaging, it was less close to the periodic
setting than \cite{Zhikov2006}.

The results in \cite{Heida2009} provide a geometrical interpretation
of subsets of the probability space and demonstrates that all periodic
quantities find their precise analogue in the stochastic case and
vice versa. Moreover, if a stationary ergodic structure is periodic,
the probability space has to be essentially the unit cell. The theory
has been applied successfully to heat transfer in polycrystals in
\cite{Heida2011}.

Due to the connection between two-scale convergence and asymptotic
expansion in the periodic setting, these mathematical results give
the basic idea to extend the asymptotic expansion to the stochastic
case. Note that it is not the intention of this article, to give a
rigorous mathematical introduction to stochastic geometry or stochastic
homogenization. Rather it is the aim of this paper to demonstrate
that it can be mathematically justified to apply results from asymptotic
expansion to stochastic settings and that there is formally no difference
in the calculations.~The most important theoretical results are cited
in the appendix. For proofs of the fundamental theorems which are
cited in this article, the reader is referred to \cite{Zhikov2006,Heida2009}
and the references therein. 

This article is organized as follows: In the next section, the basic
idea of periodic asymptotic expansion is explained and the resulting
equations for two sample problems (diffusion and porous media flow)
are given. In section \ref{sec:Stochastic-Geometry}, the concepts
of stochastic geometry will be introduced in a formal and mathematical
non rigorous way. The connections to periodic geometries are explained.
In section \ref{sec:Asymptotic-expansion} the asymptotic expansion
on stochastic geometries will be introduced and applied to thermal
diffusion, porous media flow and diffusion processes with reactions
on the microscopic boundary. In section \ref{sec:Non-periodic-models}
some simple examples of stochastic geometries will be given.

\section{Recapitulation of the Periodic Case\label{sec:Recapitulation-of-Periodic}}

In this section we will shortly recapitulate the asymptotic expansion
technique for periodic structures. It is not the intention of this
section to go into the details of calculations but to rather explain
what are the mathematical problems, the ansatz that people use to
solve these problems and what the results look like. For both examples,
precise calculations will follow in section \ref{sec:Asymptotic-expansion}
below. We will first consider the standard homogenization problem
of diffusion (following \cite{Hornung1997}) and then discuss Navier-Stokes
flow through porous media. In all calculations below and throughout
the paper, $n$ denotes the dimension of space (in a physical setting
$n=2$ or $n=3$).

\subsection{The standard problem\label{sub:The-standard-problem_per}}

Given an open and bounded set $\Q\subset\Rr^{n}$ investigate diffusion
processes (for example heat diffusion) with a source term and rapidly
oscillating coefficients which are due to a periodic microstructure.
In particular, assume that for $\Y:=[0,1[^{n}$ we have $\Y=A\cup B\cup\Gamma$
with $A$, $B$ open in $\Y$ such that $A\cap B=\emptyset$ and $\Gamma=\partial A=\partial B$
in $\Y$. We expand $\Y$ periodically to $\Rr^{n}$ and identify
$A$, $B$ and $\Gamma$ with their periodic continuation. We furthermore
denote $A^{\eps}=\eps A$, $B^{\eps}:=\eps B$ and $\Gamma^{\eps}:=\eps\Gamma$.
The diffusion constants in $A$ and $B$ are thought to be different.

The mathematical problem reads\begin{align*}
\partial_{t}u^{\eps}-\diver\left(D^{\eps}\nabla u^{\eps}\right) & =f &  & \mbox{on }]0,T]\times\Q\\
u^{\eps} & =0 &  & \mbox{on }]0,T]\times\partial\Q\\
u^{\eps}(0,\cdot) & =a(\cdot) &  & \mbox{on }\Q\mbox{ for }t=0\end{align*}
where $D^{\eps}$ is given by $D^{\eps}:=D(t,x,\frac{x}{\eps})$ and
$D$ is a $\Y$-periodic symmetric tensor with bounded entries, while
$f$ is a function on $\Q$ which may also depend on time. As an example,
we could have $D^{\eps}\equiv1$ on $A^{\eps}$ and $D^{\eps}\equiv10$
on the interior of $B^{\eps}$ with $D(t,x,y)=1+9\,\chi_{B}(y)$.

For small enough microstructures, the oscillation in $D$ may have
a minor effect on $u^{\eps}$ and the system may be described by some
averaged diffusion coefficient $D^{hom}$. It may therefore be sufficient
to solve an approximating system with this smoothed $D^{hom}$ without
the strong oscillations of $D^{\eps}$. Therefore, we investigate
$u^{\eps}$ as $\eps\rightarrow0$ by an ansatz\begin{equation}
u^{\eps}(t,x)=\sum_{i\geq0}\eps^{i}u_{i}(t,x,\frac{x}{\eps})\label{eq:AE_expand_u_eps}\end{equation}
 which is called the asymptotic expansion of $u^{\eps}$ and \[
u_{i}:\Rr_{\geq0}\times\Q\times\Y\rightarrow\Rr\]
which are $\Y$-periodic functions in the third coordinate. The gradient
operator turns into $\nabla=\nabla_{x}+\frac{1}{\eps}\nabla_{y}$
while the divergence reads $\diver=\diverx+\frac{1}{\eps}\divery$.
Altogether, the expansions above inserted into the diffusion equation
yield\begin{multline*}
\partial_{t}u_{0}-\frac{1}{\eps^{2}}\divery\left(D^{\eps}\nabla_{y}u_{0}\right)-\frac{1}{\eps}\diverx\left(D^{\eps}\nabla_{y}u_{0}\right)-\frac{1}{\eps}\divery\left(D^{\eps}\nabla_{x}u_{0}\right)-\frac{1}{\eps}\divery\left(D^{\eps}\nabla_{y}u_{1}\right)\\
-\diverx\left(D^{\eps}\nabla_{x}u_{0}\right)-\diverx\left(D^{\eps}\nabla_{y}u_{1}\right)-\divery\left(D^{\eps}\nabla_{x}u_{1}\right)-\divery\left(D^{\eps}\nabla_{y}u_{2}\right)+\mathcal{O}(\eps)=f\end{multline*}

Assuming that the terms of order $\eps^{-2}$ and $\eps^{-1}$ vanish,
it will be shown below in section \ref{sub:The-standard-problem_sto}
that this system simplifies to a single equation for $u_{0}$ \[
\partial_{t}u_{0}-\diverx\left(D^{hom}\nabla_{x}u_{0}\right)=\int_{\Y}f\]
with $D^{hom}$ defined in an appropriate way.

\subsection{Porous media flow\label{sub:Porous-media-flow-per}}

Suppose that $A^{\eps}$ is connected and consider the Navier-Stokes
equations on $\Q\cap A^{\eps}$:\begin{align*}
\partial_{t}\vel^{\eps}+\left(\vel^{\eps}\cdot\nabla\right)\vel^{\eps}-\diver\left(\nu\nabla\vel^{\eps}\right)+\nabla p^{\eps} & =f &  & \mbox{on }]0,T]\times\left(\Q\cap A^{\eps}\right)\\
\diver\vel^{\eps} & =0 &  & \mbox{on }]0,T]\times\left(\Q\cap A^{\eps}\right)\\
\vel^{\eps} & =0 &  & \mbox{on }]0,T]\times\left(\partial\Q\cup\Gamma^{\eps}\right)\\
\vel^{\eps} & =0 &  & \mbox{on }]0,T]\times B^{\eps}\\
\vel^{\eps}(0,\cdot) & =a(\cdot,\frac{\cdot}{\eps}) &  & \mbox{on }\Q\mbox{ for }t=0\end{align*}
where $f$ is an external force and $a$ is $\Y$ periodic in the
second argument. For an ansatz \begin{align*}
\vel_{i}:\Omega\times Y & \rightarrow\Rr^{3} & p_{i}:\Omega\times Y & \rightarrow\Rr^{3}\\
(x,y) & \mapsto\vel_{i}(x,y) & (x,y) & \mapsto p_{i}(x,y)\end{align*}
such that the solution $\vel^{\eps}$ and $p^{\eps}$ can be described
by \begin{align*}
\vel^{\eps} & =\sum_{i=0}^{\infty}\eps^{i}\vel_{i}(x,\frac{x}{\eps}) & p^{\eps} & =\sum_{i=0}^{\infty}\eps^{i}p_{i}(x,\frac{x}{\eps})\,,\end{align*}
the resulting set of equations is\begin{eqnarray}
-\divery(\nu\nabla_{y}\vel_{2})+\nabla_{x}p_{0}+\nabla_{y}p_{1} & = & g\nonumber \\
\divery\vel_{2} & = & 0\\
\vel_{2}(x,\cdot) & = & 0\quad\mbox{on }\partial Y_{1}\nonumber \end{eqnarray}
We will see in section \ref{sub:Porous-media-flow-sto} that for a
suitable choice of a matrix $K$, $\vel_{2}$ and $p_{0}$ fulfill
the following equation: \[
\int_{y}\vel_{2}=K\left(g-\nablax p_{0}\right)\]

which is Darcy's law.

\section{Stochastic Geometry\label{sec:Stochastic-Geometry}}

In order to develop the asymptotic expansion method for the stochastic
(non-periodic) case, it is necessary to develop a suitable mathematical
formalism. This will be the aim of the first part of this section.
As a necessary condition, we expect that the periodic case is but
a specialization of the stochastic one, which will be shown in the
second part. The basic idea of the theory described below is to replace
$\Y$ by a probability space $\Omega$. It is then necessary to identify
equivalents of $A$, $B$ and $\Gamma$ on $\Omega$ similar to the
periodic case. After that, the relations to periodic structures are
pointed out, before going on the asymptotic expansion and examples
for stochastic geometries.

\subsection{A phenomenological introduction to stochastic geometry\label{sub:phen_sto_geo}}

Instead of the periodic cell $\Y$, consider a probability space $(\Omega,\sigma,\mu)$
with the probability set $\Omega$, the sigma algebra $\sigma$ and
the probability measure $\mu$. Assume that $\Omega$ is a metric
space%
\footnote{which means there is a distance function $d:\Omega\times\Omega\rightarrow\Rr_{\geq0}$
such that $d(\omega_{1},\omega_{2})=0\Leftrightarrow\omega_{1}=\omega_{2}$,
$d(\omega_{1},\omega_{2})=d(\omega_{2},\omega_{1})$ and $d(\omega_{1},\omega_{3})\leq d(\omega_{1},\omega_{2})+d(\omega_{2},\omega_{3})$
for any $\omega_{1},\omega_{2},\omega_{3}\in\Omega$. But this is
only technical to be able to use the notion continuity.%
} and there is a family $(\tau_{x})_{x\in\Rr^{n}}$ of measurable bijective
mappings $\tau_{x}:\Omega\mapsto\Omega$ which satisfy
\begin{itemize}
\item $\tau_{x}\circ\tau_{y}=\tau_{x+y}$ 
\item $\mu(\tau_{-x}B)=\mu(B)\quad\forall x\in\Rr^{n},\,\, B\in\borelB(\Omega)$ 
\item $\mapA:\Rr^{n}\times\Omega\rightarrow\Omega\qquad(x,\omega)\mapsto\tau_{x}\omega$
is continuous 
\end{itemize}
The family $\tau_{x}$ is then called a Dynamical System. Additionally,
we claim ergodicity of $\tau_{\bullet}$, which is one of the following
two equivalent conditions\begin{equation}
\begin{split}\left[f(\omega)\stackrel{}{=}f(\tau_{x}\omega)\,\,\forall x\in\Rr^{n}\,,\, a.e.\,\,\omega\in\Omega\right] & \Rightarrow\left[f(\omega)=const\,\,\textnormal{for}\,\,\mu-a.e.\,\,\omega\in\Omega\right]\\
\left[P\left((\tau_{x}(B)\cup B)\backslash(\tau_{x}(B)\cap B)\right)=0\right] & \Rightarrow[P(B)\in\{0,1\}]\,.\end{split}
\label{eq:def_ergodicity}\end{equation}
This condition seems to be only technical, but in fact, it is crucial
for mathematical homogenization in \cite{Heida2009,Zhikov2006} and
replaces periodicity as well as the requirement that $\Y$ is simply
connected.

In order to introduce stochastic geometries from a phenomenological
point of view, assume that there is a measurable set $A\subset\Omega$
such that for the characteristic function $\chi_{A}:\Omega\rightarrow\left\{ 0,1\right\} $
holds $\chi_{A}(\tau_{\bullet}\omega):\Rr^{n}\rightarrow\left\{ 0,1\right\} $
is the characteristic function of a closed set $A(\omega)\subset\Rr^{n}$
$\mu-$almost surely in $\omega$. $A(\omega)$ is then called a random
closed set%
\footnote{Mathematically, as shown in the appendix, a random closed set is defined
via a mapping $A(\omega)$ into the set of closed sets. The existence
of sets $A,B,\Gamma\subset\Omega$ such that the above properties
are fulfilled is shown afterward. For formal calculations, it seems
more appropriate to start the other way round.%
}. Note that it possesses the property $\chi_{A(\omega)}(x+y)=\chi_{A(\tau_{x}\omega)}(y)$
which is called stationarity. 

For some of the examples below, we assume at the same time that there
is $B\subset\Omega$ such that $B(\omega)=\overline{\Rr^{n}\backslash A(\omega)}$
is the closure of the complement of $A(\omega)$. The set $\Gamma:=A\cap B$
is then also measurable and has the property that $\Gamma(\omega)=\partial A(\omega)$.
Indeed, as shown in the appendix, $\Gamma$ can be considered as an
abstract manifold, since it can be assigned a Hausdorff measure and
a normal field.

Since the space $\Omega$ is metric, there is a set of continuous
and bounded functions $C_{b}(\Omega)$. At the same time, since $A$
and $B$ are measurable subsets, it is also possible to consider continuous
functions $C_{b}(A)$ and $C_{b}(B)$ on $A$ and $B$. With help
of the dynamical system $\tau_{\bullet}$, it is possible to also
define derivatives according to \begin{equation}
\partial_{\omega,i}\, f(\omega):=\lim_{h\in\Rr\rightarrow0}\frac{f(\tau_{he_{i}}\omega)-f(\omega)}{h}\label{eq:def-partial-i}\end{equation}
where $(e_{i})_{1\leq i\leq n}$ is the canonical Basis on $\Rr^{n}$and
$f\in C_{b}(\Omega)$. The space of continuously differentiable functions
is then\[
C^{1}(\Omega):=\{f\in C(\Omega)\,\,|\,\,\textnormal{limit \eqref{eq:def-partial-i} is defined}\,\forall\omega\in\Omega\,\,\text{and }\partial_{\omega,i}f\in C(\Omega)\,\,\forall1\leq i\leq n\}\,.\]
Similar spaces may also be defined on $A$ and $B$. We may also define
the gradient $\nabla_{\omega}$ and the divergence $\diveromega$
accordingly. For any set $X$, any function $f:\Omega\rightarrow X$
and any $\omega\in\Omega$, the function $f(\tau_{\bullet}\omega):\Rr^{n}\rightarrow X$
is called a realization ($\omega$-realization) of $f$ and for any
continuous or differentiable function the realization is also continuous
or differentiable, respectively. This is due to the continuity of
$\tau$. In particular, for $f\in C_{b}(A)$ holds $f(\tau_{\bullet}\omega)\in C_{b}(A(\omega))$
and similarly for $B$ and $\Gamma$.

Finally, for a scaled random set $A^{\eps}(\omega):=\eps A(\omega)$
holds \[
\chi_{A^{\eps}(\omega)}(x)=\chi_{A(\omega)}(\frac{x}{\eps})=\chi_{A}(\tau_{\frac{x}{\eps}}\omega)\,,\]
and for a function $f\in C_{b}(A)$ holds $f(\tau_{\frac{\bullet}{\eps}}\omega)\in C_{b}(A^{\eps}(\omega))$
for all the realizations $\omega$.

\subsection{The periodic case}

It is helpful to compare the results above to the periodic case. First,
note that $\Y=[0,1[^{n}$ together with the Borel sigma algebra $\sigma_{\borelB}$
and the standard Lebesgue measure $\lebesgueL$ can be considered
as a probability space $(\Y,\sigma_{\borelB},\lebesgueL)$. For any
$x\in\Rr^{n}$ let $[x]\in\Zz^{n}$ denote the vector of integers
such that $x\in[x]+\Y$. Define the following family of mappings \begin{align*}
\tau_{x}:\Y & \rightarrow\Y\\
y & \mapsto y+x-[y+x]\end{align*}
which has all the properties we claimed for a dynamical system to
hold. For any closed set $A\subset\Y$, $A(y)$ is automatically closed
for all $y\in\Y$ and $A(0)$ is the periodic continuation of $A$.
The continuous functions on $\Y$ coincide with the $\Y$-periodic
continuous functions on $\Rr^{n}$ and the derivatives $\partial_{y,i}$
coincide with the classical derivatives $\partial_{i}$. $B$ is the
closed complement of $A$ in $\Y$ and $\Gamma$ is the boundary of
$A$ in $\Y$.

It is therefore clear, that any method which is based on the formalism
introduced above is at least applicable to the periodic setting. Section
\ref{sec:Non-periodic-models} will demonstrate that a much broader
class of geometries is covered by this approach.

\section{Asymptotic expansion on stochastic geometries\label{sec:Asymptotic-expansion}}

We are now able to introduce asymptotic expansion on stochastic geometries.
The method will be introduced via a sample calculation for the standard
homogenization problem, which is diffusion with rapidly oscillating
coefficients. This will be done using an expansion \eqref{eq:AE_expand_u_eps_stoch}
of the unknown similar to \eqref{eq:AE_expand_u_eps} in section \ref{sub:The-standard-problem_per}.
Afterward, the method will be applied to porous media flow and diffusion
with nonlinear microscopic boundary conditions. The formal calculations
are well known for the periodic case and most of them can be found
in \cite{Hornung1997} among others. Here, they will be presented
in the stochastic framework to demonstrate the similarity with the
periodic case.

\subsection{The standard homogenization problem\label{sub:The-standard-problem_sto}}

The standard problem in the stochastic setting reads\begin{align*}
\partial_{t}u^{\eps}-\diver\left(D_{\omega}^{\eps}\nabla u^{\eps}\right) & =f &  & \mbox{on }]0,T]\times\Q\\
u^{\eps} & =0 &  & \mbox{on }]0,T]\times\partial\Q\\
u^{\eps}(0,\cdot) & =a(\cdot) &  & \mbox{on }\Q\mbox{ for }t=0\end{align*}
where now $D_{\omega}^{\eps}=D(t,x,\tau_{\frac{x}{\eps}}\omega)$.
An example would be $D(\cdot,\cdot,\omega)=1$ for $\omega\in A$
and $D(\cdot,\cdot,\omega)=2$ for $\omega\in B$. 

While the original idea is to expand the functions $u^{\eps}$ asymptotically
by \eqref{eq:AE_expand_u_eps} with functions $u_{i}:\Rr_{\geq0}\times\Q\times\Y\rightarrow\Rr$,
it will now be based on a stochastic ansatz. In particular, for a
given choice of the microscopic geometry $\omega$ with the micro
structures $A^{\eps}(\omega)$, $B^{\eps}(\omega)$ and $\Gamma^{\eps}(\omega)$,
we may expand $u^{\eps}$ by \begin{equation}
u^{\eps}(t,x)=\sum_{i}\eps^{i}u_{i}(t,x,\tau_{\frac{x}{\eps}}\omega)\,.\label{eq:AE_expand_u_eps_stoch}\end{equation}
The gradient and the divergence have the expansions \begin{equation}
\nabla=\nabla_{x}+\frac{1}{\eps}\nablaomega,\quad\diver=\diverx+\frac{1}{\eps}\diveromega\label{eq:expand_grad_div_stoch}\end{equation}
Using an expansion \eqref{eq:AE_expand_u_eps_stoch} for $u^{\eps}$
with \eqref{eq:expand_grad_div_stoch} will lead to\begin{multline*}
\partial_{t}u_{0}-\frac{1}{\eps^{2}}\diveromega\left(D^{\eps}\nablaomega u_{0}\right)-\frac{1}{\eps}\diverx\left(D^{\eps}\nablaomega u_{0}\right)-\frac{1}{\eps}\diveromega\left(D^{\eps}\nabla_{x}u_{0}\right)-\frac{1}{\eps}\diveromega\left(D^{\eps}\nablaomega u_{1}\right)\\
-\diverx\left(D^{\eps}\nabla_{x}u_{0}\right)-\diverx\left(D^{\eps}\nablaomega u_{1}\right)-\diveromega\left(D^{\eps}\nabla_{x}u_{1}\right)-\diveromega\left(D^{\eps}\nablaomega u_{2}\right)+\mathcal{O}(\eps)=f\end{multline*}
which means we split up the equation in terms \[
\eps^{-2}\left(\dots\right)+\eps^{-1}\left(\dots\right)+\eps^{0}\left(\dots\right)+\mathcal{O}(\eps)=0\,.\]

Since the latter equation should hold for all $\eps$, it follows
\[
\diveromega\left(D(t,x,\omega)\nablaomega u_{0}\right)=0\,.\]
Note that in the periodic case, the latter equation implies $u_{0}(y)=const$,
while in the stochastic case, this is not clear. However, due to Gauss'
theorem \ref{stationarygauss} for dynamical systems it follows by
testing the equation with $u_{0}$: \[
\int_{\Omega}D(t,x,\omega)\left|\nablaomega u_{0}(t,x,\omega)\right|^{2}d\mu(\omega)=0\]
which is $\nablaomega u_{0}=0$. Definition \eqref{eq:def_ergodicity}$_{1}$
of ergodicity yields $u_{0}(\omega)=const$. The terms of order $\eps^{-1}$
yield \[
\diveromega\left(D^{\eps}\nabla_{x}u_{0}\right)=-\diveromega\left(D^{\eps}\nablaomega u_{1}\right)\,.\]
Now, if $\phi_{j}$ is a solution to the cell problem\[
\diveromega\left(D(t,x,\omega)\nablaomega\phi_{j}\right)=-\diveromega\left(D(t,x,\omega)e_{j}\right)\]
for $j=1,\dots,n$, the function $u_{1}$ can be expressed by \[
u_{1}(t,x,\omega)=\sum_{j=1}^{n}\phi_{j}(t,x,\omega)\partial_{j}u_{0}(t,x)\,.\]
Existence of $\phi_{j}$ with $\intomega\phi_{j}=0$ can be shown
with help of the Poincaré inequalities from the appendix and the Lax-Milgram
theorem.

Finally, the zero order terms add up to\begin{equation}
\partial_{t}u_{0}-\diveromega\left(D\nabla_{x}u_{1}-D\nablaomega u_{2}\right)-\diverx\left(D\nabla_{x}u_{0}\right)-\diverx\left(D\nablaomega u_{1}\right)=f\,.\label{eq:diff_zero_order_sto}\end{equation}
By integrating the latter equation over $\Omega$ with help of Gauss'
theorem \ref{stationarygauss}, one obtains\[
\partial_{t}u_{0}-\diverx\left(\intomega D\, d\mu\,\nabla_{x}u_{0}\right)-\intomega\diverx\left(D\nablaomega u_{1}\right)=\intomega fd\mu\,.\]
The third term on the lefthand side can be reformulated to\[
\intomega\diverx\left(D\nablaomega u_{1}\right)=\diverx\left(\intomega D\sum_{j}\nablaomega\phi_{j}\partial_{j}u_{0}\right)\]
which finally yields \[
\partial_{t}u_{0}-\diverx\left(D^{hom}\nabla_{x}u_{0}\right)=\intomega fd\mu\,.\]
Here, \[
D^{hom}=\left(D_{i,j}^{hom}\right)=\intomega\left(e_{i}D\nablaomega\phi_{j}+D_{i,j}\right)d\mu\]
is a symmetric positive definite matrix. The proof is the same as
in \cite{Hornung1997}.
\begin{rem*}
The obtained limit problem is independent on the particular choice
of the realization $\omega$ which we used for homogenization. This
is a feature of the ergodicity and stationarity and reflects our expectation
that the averaged equations should not depend on the specific geometry
but on {}``the type'' of geometry. We will see that the other examples
share this property of the limit equations.
\end{rem*}

\subsection{Porous media flow\label{sub:Porous-media-flow-sto}}

Suppose that $A^{\eps}$ is connected and consider the Navier-Stokes
equations on $\Q\cap A^{\eps}$:\begin{align*}
\partial_{t}\vel^{\eps}+\left(\vel^{\eps}\cdot\nabla\right)\vel^{\eps}-\diver\left(\nu\nabla\vel^{\eps}\right)+\nabla p^{\eps} & =f &  & \mbox{on }]0,T]\times\left(\Q\cap A^{\eps}(\omega)\right)\\
\diver\vel^{\eps} & =0 &  & \mbox{on }]0,T]\times\left(\Q\cap A^{\eps}(\omega)\right)\\
\vel^{\eps} & =0 &  & \mbox{on }]0,T]\times\left(\partial\Q\cup\Gamma^{\eps}(\omega)\right)\\
\vel^{\eps} & =0 &  & \mbox{on }]0,T]\times B^{\eps}(\omega)\\
\vel^{\eps}(0,\cdot) & =a(\cdot,\frac{\cdot}{\eps}) &  & \mbox{on }\Q\mbox{ for }t=0\end{align*}
where $f$ is an external force and $a$ is $\Y$ periodic in the
second argument. From physical investigations, we know that the flow
through a porous medium obeys Darcy's law and it is the aim of the
following calculations to demonstrate that it is possible to obtain
it as a limit problem for vanishing porescale in the stochastic setting.

For an ansatz \begin{align*}
\vel_{i}:\Rr_{\geq0}\times Q\times\Omega & \rightarrow\Rr^{3} & p_{i}:\Rr_{\geq0}\times Q\times\Omega & \rightarrow\Rr^{3}\\
(t,x,\omega) & \mapsto\vel_{i}(t,x,\omega) & (t,x,\omega) & \mapsto p_{i}(t,x,\omega)\end{align*}
such that the solution $\vel^{\eps}$ and $p^{\eps}$ can be described
by \begin{align}
\vel^{\eps} & =\sum_{i=0}^{\infty}\eps^{i}\vel_{i}(t,x,\tau_{\frac{x}{\eps}}\omega) & p^{\eps} & =\sum_{i=0}^{\infty}\eps^{i}p_{i}(t,x,\tau_{\frac{x}{\eps}}\omega)\,,\label{eqsys:Hom_AE_incomp_NS_Ansatz_eps}\end{align}
the resulting set of expanded equations is up to order $0$:\begin{subequations}\begin{multline}
\eps^{-2}\left(-\diveromega(\nu\nablaomega\vel_{0})\right)+\eps^{-1}\left(\left(\vel_{0}\cdot\nablaomega\right)\vel_{0}-\diveromega(\nu\nablaomega\vel_{1})-2\diverx(\nu\nablaomega\vel_{0})+\nablaomega p_{0}\right)\\
+\eps^{0}\left(\left(\vel_{0}\cdot\nabla_{x}\right)\vel_{0}+\left(\vel_{1}\cdot\nablaomega\right)\vel_{0}+\left(\vel_{0}\cdot\nablaomega\right)\vel_{1}\right)\\
+\eps_{0}\left(-\diveromega(\nu\nablaomega\vel_{2})-2\diverx(\nu\nabla_{x}\vel_{0})+\nabla_{x}p_{0}+\nablaomega p_{1}-g\right)+\mathcal{O}(\eps)=0\label{eq:P1_CMP_Exp_NS1_a}\end{multline}
\begin{equation}
\eps^{-1}\diveromega\vel_{0}+\diverx\vel_{0}+\diveromega\vel_{1}+\eps\left(\diverx\vel_{1}+\diveromega\vel_{2}\right)+\mathcal{O}(\eps^{2})=0\label{eq:P1_CMP_Exp_NS1_b}\end{equation}
\begin{equation}
\sum_{i}\eps^{i}\vel_{i}(x,\cdot)=0\quad\mbox{on }\partial Y_{1},\quad\sum_{i}\eps^{i}\vel_{i}(\cdot,\omega)=0\quad\mbox{on }\partial\Omega\label{eq:P1_CMP_Exp_NS1_c}\end{equation}
\end{subequations} For each power of $\eps$, a set of equations
is obtained, which has to hold independently on all the other equations
such that the whole group of equations is valid for all choices of
$\eps$.

The order of $-2$ in \eqref{eq:P1_CMP_Exp_NS1_a} together with the
order $-1$ in \eqref{eq:P1_CMP_Exp_NS1_b}, and \eqref{eq:P1_CMP_Exp_NS1_c}
yields for $\vel_{0}$\begin{eqnarray}
\eps^{-2}:\quad-\diveromega(\nu\nablaomega\vel_{0}) & = & 0\quad\mbox{on }A\nonumber \\
\diveromega\vel_{0} & = & 0\quad\mbox{on }A\label{eq:limsys_vel_0}\\
\vel_{0}(x,\cdot) & = & 0\quad\mbox{on }\Gamma\,.\nonumber \end{eqnarray}
Again, in the periodic case, it would immediately follow $\vel_{0}\equiv0$.
However, Gauss' theorem \ref{stationarygauss} again provides the
necessary framework since \[
\int_{A}\nu\left|\nablaomega\vel_{0}\right|^{2}d\mu=0\]
implies $\nablaomega\vel_{0}=0$ which is together with ergodicity
\eqref{eq:def_ergodicity} $\vel_{0}(\omega)=const$ and with \eqref{eq:limsys_vel_0}$_{3}$
finally $\vel_{0}=0$. 

The latter result together with the order $-1$ in \eqref{eq:P1_CMP_Exp_NS1_a},
order $0$ in \eqref{eq:P1_CMP_Exp_NS1_b} and \eqref{eq:P1_CMP_Exp_NS1_c}
yields for $\vel_{1}$\begin{eqnarray}
-\diveromega(\nu\nablaomega\vel_{1})+\nablaomega p_{0} & = & 0\nonumber \\
\diveromega\vel_{1} & = & 0\label{eq:limsys_vel_1}\\
\vel_{1}(x,\cdot) & = & 0\quad\mbox{on }\partial Y_{1}\nonumber \end{eqnarray}
which is again $\vel_{1}\equiv0$ due to $\int_{A}\vel_{1}\cdot\nablaomega p_{0}=0$
and \eqref{eq:limsys_vel_1}$_{2}$.

Using these results in the zero-order term in \eqref{eq:P1_CMP_Exp_NS1_a},
order 1 in \eqref{eq:P1_CMP_Exp_NS1_b} and order $2$ in \eqref{eq:P1_CMP_Exp_NS1_c},
we get \begin{equation}
\begin{split}-\diveromega(\nu\nablaomega\vel_{2})+\nabla_{x}p_{0}+\nablaomega p_{1} & =g\\
\diveromega\vel_{2} & =0\\
\vel_{2}(x,\cdot) & =0\quad\mbox{on }\Gamma\end{split}
\label{eqsys:Hom_AE_2s-Stokes}\end{equation}
Assuming that there are solutions to the problems\begin{eqnarray*}
-\diveromega(\nu\nablaomega u_{i})+\nablaomega\Pi_{i} & = & e_{i}\\
\diveromega u_{i} & = & 0\\
u_{i}(x,\cdot) & = & 0\quad\mbox{on }\Gamma\end{eqnarray*}
where $e_{i}$ is the i-th coordinate vector of $\Rr^{3}$, it is
easy to see that there is $p_{1}$ and $\vel_{2}:=\sum\left(g-\nabla_{x}p_{0}\right)_{i}u_{i}$
such that $(\vel_{2},p_{0},p_{1})$ is a solution to \eqref{eqsys:Hom_AE_2s-Stokes}.
Existence of $u_{i}$ may be again shown similarly to the previous
example. Defining a matrix $K$ by \[
K_{i,j}:=\int_{A}\nablaomega u_{i}\cdot\nablaomega u_{j}=\int_{A}u_{i}\cdot e_{j}\]
one may check that we obtain indeed Darcy's law which was expected:
\[
\int_{A}\vel_{2}=K\left(g-\nablax p_{0}\right)\]

\subsection{Diffusion with nonlinear boundary conditions}

Based on the calculations in section \ref{sub:The-standard-problem_sto},
consider the following problem\begin{align*}
\partial_{t}u^{\eps}-\diver\left(D_{\omega}^{\eps}\nabla u^{\eps}\right) & =f &  & \mbox{on }]0,T]\times\left(\Q\cap A^{\eps}(\omega)\right)\\
u^{\eps} & =0 &  & \mbox{on }]0,T]\times\partial\Q\\
\left(D_{\omega}^{\eps}\nabla u^{\eps}\right)\cdot\nu_{\Gamma^{\eps}(\omega)} & =\eps g(u^{\eps},U^{\eps}) &  & \mbox{on }]0,T]\times\left(\Gamma^{\eps}(\omega)\cap\Q\right)\\
u^{\eps}(0,\cdot) & =a(\cdot,\tau_{\frac{\bullet}{\eps}}\omega) &  & \mbox{on }\Q\cap A^{\eps}(\omega)\mbox{ for }t=0\end{align*}
combined with the nonlinear boundary problem\begin{align*}
\partial_{t}U^{\eps}+g(u^{\eps},U^{\eps}) & =0 &  & \mbox{on }]0,T]\times\left(\Gamma^{\eps}(\omega)\cap\Q\right)\\
U^{\eps}(0,\cdot) & =\tilde{U}(\cdot,\tau_{\frac{\bullet}{\eps}}\omega) &  & \mbox{on }\Gamma^{\eps}(\omega)\cap\Q\mbox{ for }t=0\,.\end{align*}
This system describes diffusion with a production or uptake due to
reactions on the microscopic boundaries. The $\eps$-factor in the
boundary condition takes into account that the microscopic surfaces
increase with a factor $\eps^{-1}$. From physical perspective, we
expect that the reactions on the walls will lead to a macroscopic
production or uptake term in the limit problem.

Using an expansion \eqref{eq:AE_expand_u_eps_stoch} for $u^{\eps}$
and a similar expansion for $U^{\eps}$, one would immediately conclude
for $w_{0}$ \begin{align*}
\partial_{t}w_{0}+g(u_{0},U_{0}) & =0 &  & \mbox{on }]0,T]\times\Q\times\Gamma\\
U_{0}(0,\cdot) & =\tilde{U}(\cdot,\omega) &  & \mbox{on }\Q\times\Gamma\mbox{ for }t=0\,.\end{align*}

For the expansion of $u^{\eps}$ one again obtains\begin{multline*}
\partial_{t}u_{0}-\frac{1}{\eps^{2}}\diveromega\left(D^{\eps}\nablaomega u_{0}\right)-\frac{1}{\eps}\diverx\left(D^{\eps}\nablaomega u_{0}\right)-\frac{1}{\eps}\diveromega\left(D^{\eps}\nabla_{x}u_{0}\right)-\frac{1}{\eps}\diveromega\left(D^{\eps}\nablaomega u_{1}\right)\\
-\diverx\left(D^{\eps}\nabla_{x}u_{0}\right)-\diverx\left(D^{\eps}\nablaomega u_{1}\right)-\diveromega\left(D^{\eps}\nabla_{x}u_{1}\right)-\diveromega\left(D^{\eps}\nablaomega u_{2}\right)+\mathcal{O}(\eps)=f\end{multline*}
together with the following boundary conditions on $\Gamma$:\begin{equation}
\frac{1}{\eps^{2}}\left(D\nablaomega u_{0}\right)\cdot\nu_{\Gamma}+\frac{1}{\eps}D\left(\nabla_{x}u_{0}+\nablaomega u_{1}\right)\cdot\nu_{\Gamma}+D\left(\nabla_{x}u_{1}+\nablaomega u_{2}\right)\cdot\nu_{\Gamma}-g(u_{0},U_{0})=0\,.\label{eq:limit_BC_sto}\end{equation}
It is again possible to obtain with slightly modified argumentation
in the partial integrations $u_{0}(\omega)=const$ as well as \[
u_{1}(t,x,\omega)=\sum_{j=1}^{n}\phi_{j}(t,x,\omega)\partial_{j}u_{0}(t,x)\,,\]
where now the $\phi_{j}$ are solutions to \begin{align*}
\diveromega\left(D(t,x,\omega)\nablaomega\phi_{j}\right) & =-\diveromega\left(D(t,x,\omega)e_{j}\right) &  & \mbox{on }A\\
\left(D\nablaomega\phi_{j}+De_{j}\right)\cdot\nu_{\Gamma} & =0 &  & \mbox{on }\Gamma\,.\end{align*}

The zero order terms in \eqref{eq:diff_zero_order_sto} are now integrated
over $A$ which yields together with the zero order boundary condition
in \eqref{eq:limit_BC_sto}:\[
\partial_{t}u_{0}-\diverx\left(D^{hom}\nabla_{x}u_{0}\right)-\int_{\Gamma}g(u_{0},U_{0})d\mu=\intomega fd\mu\,.\]
where now \[
D^{hom}=\left(D_{i,j}^{hom}\right)=\int_{A}\left(e_{i}D\nablaomega\phi_{j}+D_{i,j}\right)d\mu\,.\]
We therefore find the nice property that the macroscopic unknown $u_{0}$
is produced or consumed by microscopic reactions on the surfaces,
which was expected.

\section{Non-periodic models captured by stochastic geometries\label{sec:Non-periodic-models}}

We will now describe some basic modeling tools for stochastic geometries
and give some simple but interesting models for different applications
in microstructures. Note that for many real world applications, it
is up to now not possible to give a model for the natural geometries.
For example natural soil with roots, wormholes and other obstacles
is currently out of reach. Therefore, the models given here can only
be considered as very rough models for porous media and other applications.
Also, we will restrict on the most simple models. For more models
in stochastic geometry, refer to Stoyan, Kendall and Mecke \cite{Stoyan2008}.

\subsection{Point processes}

A point process (PP) is a measurable mapping $A:\Omega\rightarrow\left(\Rr^{n}\right)^{\Nn}$.
Let $\mathP(N,Q,p)$ denote the probability to find $N$ points in
the bounded and open set $Q\subset\Rr^{n}$ for the point process
$p$. For some open set $V\subset\Rr^{n}$ with $V\ni0$, define the
intensity $\lambda_{p}$ by\[
\lambda_{p}:=\lim_{t\rightarrow\infty}\frac{1}{t^{n}\left|V\right|}E\left(\mathP(\cdot,tV,p)\right)\,,\]
where $E(\mathcal{P}(\dots))$ is the expectation value of $N$. According
to \cite{Stoyan2008} a PP is stationary if its characteristic is
invariant under translation, i.e. $\mathP(N,Q,p)=\mathP(N,Q+x,p)$
for all $x\in\Rr^{n}$. From a stochastic point processes, one may
construct higher dimensional structures in a deterministic way. Such
structures would still be stationary random closed set. We will come
to that point below. 

A prominent example of a PP is the so called stationary Poisson point
process which has the property that \[
\mathP(N,Q,p)=\frac{\lambda_{p}^{N}\left|Q\right|^{N}}{N!}\exp\left(-\lambda_{p}\left|Q\right|\right)\]

An other useful class of PP are so called hard core PP: From a given
random PP any point is erased if its nearest neighbor has a distance
less than a certain value $d$.

\subsection{Voronoi-Tessellations and Delaunay-Diagram}

\begin{figure}
\includegraphics[bb=2bp 2bp 15bp 10bp,clip,width=7cm]{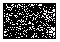}\hspace{0.5cm}\includegraphics[bb=2bp 2bp 15bp 10bp,clip,width=7cm]{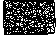}

\caption{Left: Poisson-Voronoi tessellation with its underlying Poisson point
process; Right: The corresponding Delaunay tessellation. The point
processes and tessellations where constructed using the {}``Stochastic
Geometry 4.1'' software, developed at the TU Bergakademie Freiberg,
Institut für Stochastik.\label{fig:PV-Tessellation}}

\end{figure}
Starting from a PP, one may construct the so called Voronoi-tessellation
as the set of all points in $\Rr^{n}$, whose nearest two neighbors
of the PP have the same distance. This can be used for example as
a model of polycrystals\cite{Heida2009}: The point process represents
the set of crystallization nuclei who start to grow at the same time
with the same speed and who stop growing the moment they hit each
other.

A generalization of this model which models crystal growth starting
at different times with different speed is the so called Johnson-Mehl-tessellation.

As a dual of the Voronoi-tessellation, one can consider the Delaunay-diagram
which connects all points of the PP who share a common border in the
Voronoi-tessellation. Connecting the points with cylinders (pipes)
instead of lines, one would already get a model for a porous medium.
Examples of both are shown in figure \ref{fig:PV-Tessellation}

\subsection{Ball- and Grain-models}

\begin{figure}

\includegraphics[bb=2bp 2bp 15bp 10bp,clip,width=7cm]{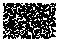}\caption{A ball process based on a Mattern hard core point process of type
2 (generated with Stochastic Geometry 4.1)\label{fig:ball-process}}

\end{figure}
Based on a hard core PP, assign to each point of the PP a ball with
a fixed radius (Figure \ref{fig:ball-process}). This model could
be used to model a porous medium in 2D. However, in 3D we expect the
grains of the porous medium matrix to touch some neighbors and that
none of them is free without touching any neighbor. To achieve such
a geometry, consider the following construction:

From a Voronoi-tessellation consider for each point $P_{c}$ of the
PP the set of points $P_{b,i}$ where the corresponding Delaunay-diagram
hits the grain boundaries. We may then interpolate these points by
a smooth manifold to mark the surface of a grain. As an example, define\[
d(P_{c},x):=\sum_{i}\frac{(x-P_{c})^{2}}{(P_{b,,i}-P_{c})^{2}}\prod_{j\not=i}\frac{(x-P_{b,,j})^{2}}{(P_{b,i}-P_{b,,j})^{2}},\quad G(P_{c}):=\left\{ x\in\Rr^{n}:d(P_{c},x)\leq1\right\} \]
The sets $G(\cdot)$ could then model the grains of sand and the complement
of the $G$ would be the porous medium.

\section{Conclusion}

We saw that asymptotic expansion can also be applied to stochastic
geometries if they fulfill certain conditions, in particular stationarity
and ergodicity. The method was applied to diffusion with and without
nonlinear microscopic boundary conditions and to porous media flow.
Some examples for stochastic geometries where given for solid and
porous microstructures. Note that we did not treat homogenization
of problems $\partial_{t}u^{\eps}-\diver\left(\eps^{2}D^{\eps}\nabla u^{\eps}\right)=f$
with $\eps^{2}$-scaled diffusion. However, such problems can be treated
even easier than the example calculation above with results of the
form $\partial_{t}u_{0}-\diveromega\left(D\nablaomega u_{0}\right)=f$.

The results of this article can be considered as one more step towards
understanding homogenization in non-periodic heterogeneous media.
A big advantage of the presented approach is, that formal calculations
require only few new theory. The only formal difference in the calculation
is, that $\Y$ has to be changed into $\Omega$ and $y$ into $\omega$.
Although the mathematical theory behind the rigorous calculations
gets much more complex, the method's user is not bothered by that.
Indeed, besides some critical points in the calculations which were
pointed out, one may not even think of the fact that one is doing
stochastic calculations. The limit equations are independent on the
particular realization which was used for homogenization. This is
a feature of ergodicity and stationarity of the geometries.

A major issue for further investigations is the search for suitable
models for geometries of natural heterogeneous media. Also, from the
mathematical point of view, it would be nice to develop a {}``stochastic
unfolding'' corresponding to the periodic unfolding as a further
theoretical foundation of the presented asymptotic expansion. Finally
remark, that the intention of this paper was not to give a rigorous
introduction to stochastic homogenization or the mathematics behind
it but only to demonstrate that for non-mathematicians, switching
from periodicity to stationary ergodic geometries can be easily achieved
and would not bother the calculations or the results. 

\appendix

\section{Stochastic Geometries}

This section follows \cite{Heida2009} to introduce some basic mathematical
concepts on random geometries. In particular, random closed sets and
random measures will be defined rigorously. The connection to the
periodic case was explained in detail in \cite{Heida2009} and is
based on similar ideas as the connections between periodic and stochastic
asymptotic expansion. It will be shown that the assumptions on $A$,
$B$ and $\Gamma$ in section \ref{sub:phen_sto_geo} are reasonable
and can be made w.l.o.g.. For more information on random closed sets,
refer to Matheron \cite{Matheron1975} or Molchanov \cite{Molchanov2005}.

Let $\closedsets(\Rr^{n})$ denote the set of all closed sets in $\Rr^{n}$.
Then for $K\subset\Rr^{n}$ compact and $V\subset\Rr^{n}$ the following
sets can be defined: \begin{eqnarray}
\closedsets_{V}:= & \left\{ F\in\closedsets(\Rr^{n})\quad|\quad F\cap V\not=\emptyset\right\}  & V\subset\Rr^{n}\quad\textnormal{open set}\label{eq:fell-sets_V}\\
\closedsets^{K}:= & \left\{ F\in\closedsets(\Rr^{n})\quad|\quad F\cap K=\emptyset\right\}  & K\subset\Rr^{n}\quad\textnormal{compact set}\end{eqnarray}
The Fell topology on $\closedsets(\Rr^{n})$ is created by the sets
$\closedsets_{V}$, $\closedsets^{K}$ for all open V and compact
K and according to \cite{Matheron1975}, $(\closedsets(\Rr^{n}),\ttopology)$
is compact, Hausdorff and separable. The \emph{Matheron-$\sigma$-field
$\sigma_{\closedsets}$} is the Borel-$\sigma$-algebra created by
the Fell-topology.

For a probability space $(\Omega,\sigma,\mu)$, a \emph{Random Closed
Set (RACS)} is a measurable mapping \[
\tilde{A}:(\Omega,\sigma,\mu)\longrightarrow(\closedsets,\sigma_{\closedsets})\,.\]
It is the aim of this short introduction to demonstrate that such
RACS have the nice properties which we assumed on $A$, $B$ and $\Gamma$
from section \ref{sub:phen_sto_geo}. Remark that up to now, a RACS
is a mapping from a probability space into the set of closed subsets
of $\Rr^{n}$. 

In what follows, $\baireM(\Rr^{n})$ respectively $\baireM$ denotes
the set of all locally finite Borel measures on $\Rr^{n}$. In Particular,
$\lebesgueL$ denotes the Lebesgue measure and $\hausdorffH^{m}$
the $m$-dimensional Hausdorff measure. The smallest topology such
that \[
\baireM\rightarrow\Rr,\quad\tilde{\mu}\longmapsto\int f\, d\tilde{\mu}\]
 is continuous for all $f\in C_{0}(\Rr^{n})$ is in general called
the Vague-topology $\vaguetopology$ on $\baireM(\Rr^{n})$. The Borel-$\sigma$-field
of this topology is denoted by $\sigma_{\mathcal{V}}$ respectively
by $\borelB(\baireM)$.

\begin{theorem}\cite{Daley1988}\label{lemmameasureableset} \label{thbaireMisseparablemetric}
$(\baireM,\vaguetopology)$ is a separable metric space. The $\sigma$-algebra
$\sigma_{\mathcal{V}}$ on $\baireM$ created by the Vague topology
$\vaguetopology$ is the smallest one, such that $\mu\mapsto\mu(B)$
is measurable for every bounded and measurable set $B\subset\Rr^{n}$.
\end{theorem}

A \emph{random measure} on $\Rr^{n}$ is a measurable mapping $(\Omega,\sigma,\mu)\rightarrow(\baireM,\sigma_{\vagueV})$,
$\omega\mapsto\mu_{\omega}$, where $(\Omega,\sigma,\mu)$ is a probability
space. Random closed sets and the random measures are related due
to the following Lemma, which implies that a random closed set always
induces a random measure. 

\begin{lemma}(\cite{Zaehle1982} Theorem 2.1.3 resp. Corollary 2.1.5)
\label{lemmazaehlerandommeasure} 

Let $\closedsets_{m}\subset\closedsets$ be the space of closed m-dimensional
sub manifolds of $\Rr^{n}$ such that the corresponding Hausdorff
measure is locally finite. Then, the $\sigma$-algebra $\sigma_{\closedsets}\cap\closedsets_{m}$
is the smallest such that \[
M_{B}^{m}:\closedsets_{m}\rightarrow\Rr\quad\Gamma\mapsto{\mathcal{H}}^{m}(\Gamma\cap B)\]
 is measurable for every measurable and bounded $B\subset\Rr^{n}$.
\end{lemma}

It was part of the argumentation in \cite{Heida2009} that due to
Lemma \ref{lemmazaehlerandommeasure} for any initial random closed
manifold $A:(\Omega,\sigma,\mu)\rightarrow\closedsets_{m}$ the assumptions
$\Omega\subset\baireM$ and $\sigma=\borelB(\baireM)\cap\Omega$ can
be made w.l.o.g.. We introduced dynamical systems in section \ref{sub:phen_sto_geo}
and also defined ergodicity. However, one normally assumes only measurability
of $\tau$ but the continuity is a rather direct consequence of $\Omega\subset\baireM$
\cite{Heida2009}. 

A random measure is said to be \emph{stationary} if for every $\omega\in\Omega$:
$T(x)\mu_{\omega}=\mu_{\tau_{x}\omega}$ with \[
T(x)\mu_{\omega}(B):=\mu_{\omega}(B+x)\quad\forall B\in\borelB(\Rr^{n})\]
and a RACS $A$ is called stationary if $\chi_{A(\tau_{y}\omega)}(x)=\chi_{A(\omega)}(x+y)$
where $\chi_{A(\omega)}$ is the characteristic function of $A(\omega)$.
This is slightly different introduction of stationarity than in section
\ref{sub:phen_sto_geo}. However, we will see below that these definitions
of stationarity and ergodicity already guaranty that RACS have the
properties which we claimed in section \ref{sub:phen_sto_geo}. First,
it is necessary to show that there is an equivalent of a Hausdorff-measure
on $\Gamma$. This Hausdorff measure will be $\mugammapalm$, stated
by the following

\begin{theorem}\label{theoremmecke}(Mecke \cite{Mecke1967,Daley1988}:
Existence of Palm measure)

Let $\lebesgueL$ be the Lebesgue-measure on $\Rr^{n}$ with $dx:=d\lebesgueL(x)$
and $(\Omega,\sigma,\mu)$ as above. Then there exists a unique measure
$\mupalm$ on $\Omega$ such that \begin{equation}
\intomega\intrn f(x,\tau_{x}\omega)\, d\muomega(x)d\mu(\omega)=\intrn\intomega f(x,\omega)\, d\mupalm(\omega)dx\label{eq:palm_equation}\end{equation}
 for all $\borelB(\Rr^{n})\times\borelB(\Omega)$-measurable non negative
functions and all $\mupalm\times\lebesgueL$- integrable functions.
\end{theorem}

$\mupalm$ as in Theorem \ref{theoremmecke} is called the \emph{Palm
measure} of $\mu_{\omega}$. Note that $\mu$ is the Palm measure
of $\lebesgueL$. For a random closed manifold $\Gamma:\,(\Omega,\sigma,\mu)\rightarrow\closedsets_{m}\quad\omega\mapsto\Gamma(\omega)$
the random Hausdorff measure will be denoted by $\omega\mapsto\mu_{\Gamma(\omega)}$
and the Palm measure by $\mugammapalm$.

\begin{theorem}(Ergodic Theorem \cite{Daley1988})\label{ergodictheorem}\\
 Let the dynamical System $\tau_{x}$ be ergodic and assume that
the stationary random measure $\mu_{\omega}$ has finite intensity.
Then \begin{equation}
\lim_{t\rightarrow\infty}\frac{1}{t^{n}\lebesgueL(A)}\int_{tA}g(\tau_{x}\omega)d\muomega(x)=\int_{\Omega}g(\omega)d\mupalm(\omega)\end{equation}
 for $\mu$ almost surely in $\omega$ for all bounded Borel sets
$A\supset\{|x|<1\}$ and all $g\in L^{1}(\Omega,\mupalm)$ \end{theorem}

For $\mugammaomegaeps(B):=\eps^{n}\mugammaomega(\eps^{-1}B)$ the
latter theorem yields after a transformation of variables\begin{equation}
\lim_{\eps\rightarrow0}\int_{Q}f(x,\tau_{\frac{x}{\eps}}\omega)d\mugammaomegaeps=\int_{Q}\int_{\Omega}f(x,\omega)d\mugammapalm dx\,.\label{eq:ergodic_theorem}\end{equation}
for $f=\phi(x)\psi(\omega)$ with $\phi\in C_{0}^{\infty}(Q)$ and
$\psi\in C_{b}(\Omega)$ and therefore also for $f\in C_{0}^{\infty}(Q;C_{b}(\Omega))$.
This is a first indicator why we may consider $\mugammapalm$ as a
Hausdorff measure. It will be more striking by the following

\begin{lemma}\label{gammahasmeasurezero}\cite{Heida2009} There
is a measurable set, also denoted by $\Gamma\subset\Omega$, with
$\chi_{\Gamma(\omega)}(x)=\chi_{\Gamma}(\tau_{x}\omega)$ for $\lebesgueL+\mugammaomega$-almost
every $x$ for $\mu$-almost every $\omega$. Furthermore $\mu(\Gamma)=0$
and $\mugammapalm(\Omega\backslash\Gamma)=0$. \end{lemma}

The same proof also provides such a characteristic function for a
$n$-dimensional random closed set $\tilde{A}$. $\tilde{\Gamma}:=\partial\tilde{A}$
is also a random closed set, which can be verified using $\closedsets_{V}$
from \eqref{eq:fell-sets_V} with $V=\tilde{A}\backslash\partial\tilde{A}$.
In case $\tilde{\Gamma}$ is regular enough, there are subsets $A,\Gamma\subset\Omega$
such that $\chi_{\tilde{A}(\omega)}(x)=\chi_{A}(\tau_{x}\omega)$
and $\chi_{\tilde{\Gamma}(\omega)}(x)=\chi_{\Gamma}(\tau_{x}\omega)$.
This is equally possible for $\tilde{B}:=\left(\Rr^{n}\backslash\tilde{A}\right)\cup\tilde{\Gamma}$.
Thus, the assumptions made on $A$, $B$, and $\Gamma$ made in section
\ref{sub:phen_sto_geo} are now justified. Note that it is also possible
to define $\nu_{\Gamma}:\Gamma\rightarrow\Rr^{n}$ such that $\nu_{\Gamma(\omega)}(x)=\nu_{\Gamma}(\tau_{x}\omega)$
and $\nu_{\Gamma}$ thus can be considered as normal field on $\Gamma$. 

Note that in section \ref{sub:phen_sto_geo}, we started with $A,B,\Gamma\subset\Omega$
and defined $A(\omega)$, $B(\omega)$ and $\Gamma(\omega)$ to avoid
these rather technical preliminaries. We conclude the mathematical
part by some remarks on function spaces on $\Omega$ as well as on
Gauss' theorem and Poincaré inequalities.

Continuity and differentiability of functions were introduced in section
\ref{sub:phen_sto_geo}. It is of course also possible to define arbitrarily
high differentiability $C_{b}^{k}(\Omega)$. Since $\Omega$ is a
separable metric space equipped with a measure $\mu$, it is possible
to define \[
L^{p}(\Omega,\sigma,\mu):=\left\{ f:\Omega\rightarrow\Rr\,:\, f\mbox{ is }\sigma-\mbox{measurable},\,\,\int\left|f\right|^{p}d\mu<\infty\right\} \]
 These $L^{p}$-spaces have good properties, in particular they are
separable and for any finite measure $\mu$ on $(\Omega,\borelB(\Omega))$
exists a countable dense set of $C_{b}^{\infty}(\Omega)$-functions
in $L^{p}\spaceomega$ ($1\leq p<\infty$, $\mu$ \sigmafinite)\cite{Heida2009}.

It is of course possible to define various kinds of Sobolev spaces
on $\Omega$. The reader is referred to \cite{Zhikov1994,Zhikov2006,Heida2009}.

Finally, in order to justify some of the calculations in section \ref{sec:Asymptotic-expansion},
it is vital to proof a Gauss-like theorem on stochastic spaces as
well as some Poincaré inequalities. 

\begin{lemma}\cite{Heida2009}\label{stationarygauss} For all $\psi\in C_{b}^{1}(\Omega)$
and $\phi\in C_{b}^{1}(\Omega)^{n}$ holds: \begin{equation}
\int_{\Omega}\psi\nabla\phi d\mu=-\int_{\Omega}(\nabla\psi)\phi d\mu\label{eq:Gauss}\end{equation}
 \end{lemma} Since the technique is very often used, we give the
abbreviated proof from \cite{Heida2009}:\\
 Define $Q_{m}:=[-m,m]^{n}$. Then, since $\phi$ and $\psi$ are
bounded, the Ergodic Theorem \ref{ergodictheorem} leads to: \begin{eqnarray*}
\int_{\Omega}\psi\nabla\phi d\mu & = & \lim_{m\rightarrow\infty}\frac{1}{(2m)^{n}}\int_{Q_{m}}\psi(\tau_{x}\omega)\nabla\phi(\tau_{x}\omega)dx\\
 & = & \lim_{m\rightarrow\infty}\frac{1}{(2m)^{n}}\left(-\int_{Q_{m}}(\nabla\psi)(\tau_{x}\omega)\phi(\tau_{x}\omega)dx\right.\\
 &  & \phantom{\lim_{m\rightarrow\infty}\frac{1}{(2m)^{n}}\Big(}\left.+\int_{\partial Q_{m}}\nu_{Q_{m}}(x)(\phi\psi)(\tau_{x}\omega)d{\mathcal{H}}^{n-1}(x)\right)\\
 & = & =\lim_{m\rightarrow\infty}\frac{1}{(2m)^{n}}\left(-\int_{Q_{m}}(\nabla\psi)(\tau_{x}\omega)\phi(\tau_{x}\omega)dx+\mathcal{O}(m^{n-1})\right)\\
 & = & -\int_{\Omega}(\nabla\psi)\phi d\mu\end{eqnarray*}
where the integral over the boundary vanishes due to the fact that
the surface of $Q_{m}$ grows with $(2m)^{n-1}$.QED

Note that \eqref{eq:Gauss} would also hold for an integral over $A$
if $\psi=0$ on $\Gamma$ or $\nabla\psi\cdot\nu_{\Gamma}=0$ on $\Gamma$
.

From the corresponding Poincaré inequalities on $\Rr^{n}$, one may
conclude in the same way\begin{eqnarray*}
\intomega\left(\psi^{2}+\left|\nabla\psi\right|^{2}\right) & \leq & C\left(\intomega\psi+\intomega\left|\nabla\psi\right|^{2}\right)\quad\forall\psi\in C_{b}^{1}(\Omega)\\
\int_{A}\left(\psi^{2}+\left|\nabla\psi\right|^{2}\right) & \leq & C\int_{A}\left|\nabla\psi\right|^{2}\quad\forall\psi\in C_{b}^{1}(A)\mbox{ with }\psi=0\mbox{ on }\Gamma\end{eqnarray*}
where we have to assume that the constant $C$ for the realizations
is independent on $\omega$:\[
\int_{Q_{m}\cap A(\omega)}\left(\psi^{2}(\tau_{x}\omega)+\left|\nabla\psi(\tau_{x}\omega)\right|^{2}\right)\leq C\int_{Q_{m}\cap A(\omega)}\left|\nabla\psi(\tau_{x}\omega)\right|^{2}+\mathcal{O}(m^{n-1})\]
Once such inequalities are established for $C_{b}^{1}$-functions,
it is easy to expand them to corresponding Sobolev spaces on $\Omega$.
(See \cite{Heida2009,Heida2011} for more complicated examples)

\bibliographystyle{plain}
\bibliography{StoAsEx}

\end{document}